# Glocalizing Generative AI in Education for the Global South: The Design Case of 21st Century Teacher Educator AI for Ghana

Matthew Nyaaba

Department of Educational Theory and Practice, University of Georgia
AI4STEM Education Center, University of Georgia

**Abstract**

This study presents the design and development of the *21st Century Teacher Educator for Ghana GPT*, a customized Generative AI (GenAI) tool developed using OpenAI's Retrieval-Augmented Generation (RAG) framework and the Interactive Semi-Automated Prompting Strategy (ISA). Anchored in a Glocalized approach, this AI tool is designed to support pre-service teachers (PSTs) in Ghana by integrating localized linguistic, cultural, and curricular knowledge into the broader context of global AI ethics and responsible use. The GPT model draws upon structured, preloaded datasets—including Ghana's National Teacher Education Curriculum Framework (NTECF), UNESCO's (2023) human-centered AI guidelines, and culturally responsive education frameworks, to provide curriculum-aligned, linguistically adaptive, and pedagogically grounded support. The ISA assist PSTs to identify and select their institution, year, and semester, the system delivers context-specific learning resources, including lecture summaries, practicum support, assessment practice, and action research guidance, thereby fostering self-directed learning and critical thinking among PSTs. The design foregrounds equity and cultural relevance by incorporating multiple frameworks: the Culture and Context-Aware Framework, GenAI-CRSciA, Mitigating Neocolonialism in GenAI, and national curriculum standards, ensuring fidelity to Ghanaian educational values and responsiveness to diverse learner needs. Pilot implementation revealed the GPT's unique capabilities in language adaptation and localization, offering bilingual instruction in English and indigenous Ghanaian languages such as *Twi, Dagbani, Mampruli, and Dagaare*, with culturally grounded examples to enhance conceptual clarity. Notable generating practice assessments aligned with course objectives. One key challenge was the occasional hallucinations, particularly in underrepresented local languages. Also, subscription to premium will be a challenge for PSTs full access. This design case contributes to emerging discourses on Glocalized GenAI in education and calls for collaboration with OpenAI NextGen to expand full access and empirical research to assess PSTs' engagement and optimize multilingual GenAI systems across varied African contexts.

*Keywords: Generative AI, teacher education, pre-service teachers, action research, AI in education, context-aware AI, Ghanaian curriculum.*

## 1.0 Introduction

The integration of Generative Artificial Intelligence (GenAI) in teacher education has become increasingly prevalent, with many pre-service teachers (PSTs) leveraging GenAI tools to support both learning and research (Nyaaba, 2024). As AI adoption grows globally, concerns have emerged regarding the ability of GenAI systems to generate context-specific, pedagogically appropriate, and culturally relevant responses—especially in the Global South, where educational data is often underrepresented in mainstream training datasets (Nyaaba, Shi, et al., 2024). These limitations emphasize the urgent need for customized GenAI tools that address the specific challenges PSTs face in contexts like Ghana. The accuracy, trustworthiness, and contextual alignment of GenAI-generated content are critical to determining whether PSTs can confidently use these technologies for academic growth and professional development.

In Ghana, PSTs follow a structured four-year Bachelor of Education (B.Ed.) curriculum that includes foundational coursework, pedagogical training, practicum experience, and a final action research project. Recent educational reforms by the National Council for Curriculum and Assessment (NaCCA) have integrated digital competencies, foundational AI literacy, and computational thinking into teacher training programs (GMOE, 2018; NaCCA, 2020). These reforms are regulated by the National Teaching Council (NTC) and aligned with the National Teacher Education Curriculum Framework (NTECF) to standardize quality across teacher preparation institutions. However, research indicates that many teacher educators struggle to teach these evolving standards effectively, often resulting in curriculum infidelity (Nyaaba et al., 2024). This points to an urgent need for technology-enabled professional support systems that can bridge curriculum gaps.

GenAI has shown potential to promote learner agency, autonomy, and collaboration, particularly among PSTs who are increasingly adopting AI tools for self-directed learning. In contrast, teacher educators often remain at the "observer" stage of the GenAI adoption continuum (Zhai, 2025), which includes the stages of observer → adopter → collaborator → innovator (Nyaaba et al., 2024). This imbalance necessitates intentional design interventions to elevate both PST and educator engagement with AI.

This design case responds to five key needs:

1. Enhancing instructional quality and learner engagement to promote curriculum fidelity, a persistent challenge within Ghana's teacher education
2. Enhancing PSTs' use of GenAI as a smart learning companion that fosters prompt-based collaboration and inquiry aligned with the NTECF curriculum.
3. Supporting teacher educators to move beyond observation toward active adoption, collaboration, and innovation with GenAI technologies.
4. Addressing persistent issues of cultural bias, contextual inaccuracy, and ethical limitations found in mainstream GenAI applications.
5. Fostering 21st-century competencies among PSTs, including digital literacy, critical thinking, and culturally responsive pedagogical practices.

Importantly, this design also adheres to the UNESCO (2023) framework for the ethical use of GenAI in education, which emphasizes human agency, pedagogically responsible use, and the importance of maintaining human–AI co-regulation in learning environments. The design aligns with global guidance that educators and researchers should prioritize human-centered, responsible AI interaction when determining how and when to use GenAI tools in teaching and learning.

We utilized OpenAI's infrastructure to create a scalable and glocalized GPT. Configuration materials included open-source Ghanaian curriculum documents, teaching syllabi, lecture notes, sample assessments, and culturally responsive teaching frameworks. We embedded Interactive Semi-Automated (ISA) prompting strategies, which reduce the burden of crafting effective prompts while facilitating proactive, human-guided dialogue between PSTs and the GPT (Nyaaba & Zhai, 2025). This design enhances curriculum fidelity and cultural inclusion and contributes to mitigating the data underrepresentation of the Global South in GenAI systems, providing a practical model for glocalized and responsible GenAI integration in teacher education.

## 2. Theoretical Framework

As GenAI systems increasingly influence how knowledge is accessed, processed, and shared in educational contexts, the urgent need for Glocalized GenAI design has come into sharper focus. Glocalized GenAI refers to artificial intelligence tools that are globally aware yet locally grounded, technologically advanced but culturally and pedagogically aligned with the diverse realities of learners and educators. The design of such systems must go beyond technical considerations to embrace ethical, cultural, curricular, and decolonial frameworks. Drawing from recent scholarships and field-tested innovation, this section proposes a *multi-layered theoretical foundation* for the development and use of glocalized GenAI tools across interdisciplinary educational settings.

At the core of this foundation is the *UNESCO (2023) framework on co-designing GenAI for education*, which asserts that GenAI must enhance, not undermine, *human agency, teacher professionalism, and pedagogical sovereignty*. This principle is essential for countering tendencies toward automation that bypass teacher expertise or reduce complex human processes to algorithmic routines. Co-designing AI with educators, students, and local communities ensures that the resulting tools align with shared learning goals and cultural values. Building on this, the *Culture and Context-Aware Framework* (Nyaaba & Zhai, 2025) offers a structure for embedding *cultural competence* and *pedagogical relevance* into GenAI systems. GenAI integrates insights from *Gay's (2000)* model of culturally responsive teaching, *Ladson-Billings' (1994)* work on culturally relevant pedagogy, and *contextual intelligence theory* (Schlitt & Weiser, 1994). These elements encourage the inclusion of local languages, sociocultural references, and curriculum fidelity in GenAI outputs, features that are crucial for learner engagement and equity in any setting, from teacher education to STEM, social studies, or health sciences.

To address the *assessment and feedback dimension* of GenAI-enhanced learning, the *GenAI-CRSciA framework* (Nyaaba et al., 2024) adds a layer of fairness and representation. Originally developed for culturally responsive science assessment, the GenAI-*CRSciA framework* can be adapted across disciplines to ensure that GenAI-generated evaluations are inclusive of diverse linguistic, racial, spiritual, and community-based perspectives. It promotes the design of *human-centered feedback systems* that affirm identity and avoid the replication of dominant norms. The *framework for mitigating GenAI neocolonialism* (Nyaaba et al., 2024) provides design strategies to prevent the reproduction of Western-centric data hierarchies in educational GenAI. It introduces two key methodologies: *Liberatory Design Methods (LDM)*, which empower developers to elevate non-Western epistemologies and view marginalization as a source of innovation (Calzati, 2021; Harrington & Piper, 2018), and *Foresight by Design (FBD)*, which enables anticipatory thinking and harm reduction in AI development by incorporating pluralistic futures and ethical foresight (Buehring & Liedtka, 2018). These approaches advocate for *decentralized AI development, local design hubs, and contextualized prompt engineering*, strategies that are not only ethical but vital for ensuring that GenAI technologies do not reinforce global inequities in knowledge representation and power.

Finally, frameworks such as Ghana's *National Teacher Education Curriculum Framework (NTECF)* serve as a model for how *local curricular standards* can be operationalized in GenAI tools. The NTECF emphasizes interactive, inclusive, and practice-based learning aligned with teacher standards. Similar structures exist across disciplines, be it engineering ethics, nursing codes of practice, or early childhood development benchmarks. These can serve as *disciplinary anchors* in glocalized AI development, ensuring that tools support, not sideline, professional norms and learning trajectories. These frameworks support the development of GenAI tools that are: *Ethically aligned, Culturally grounded, Pedagogically relevant, Technologically innovative, Locally inclusive and globally conscious.* This theoretical foundation can be applied across disciplines—from teacher education to public health, from law to agriculture—to build GenAI systems that honor context, empower users, and foster *interconnected, inclusive learning ecosystems.* In doing so, glocalized GenAI becomes not only a tool for access but a *vehicle for equity, knowledge justice, and transformative education*.

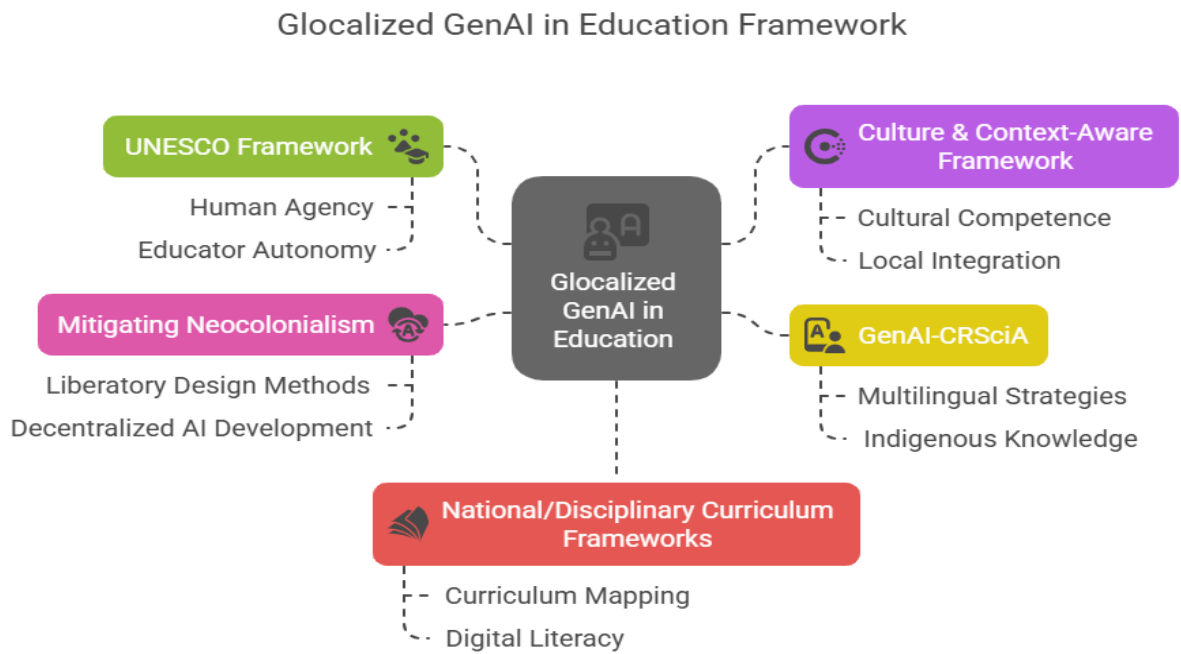

*Figure 1: Glocalized Generative Ai Education Framework*

## 2.1 Potentials and Limitations of Customized GPTs

Recent studies show the potential benefits of custom GPT in education. Kabir et al. (2024) investigated the potential of custom GPTs in scientific writing by developing two AI-powered tools, the Neurosurgical Research Paper Writer and the Medi Research Assistant. These models were refined iteratively using OpenAI's GPT Builder, incorporating specific instructions and feedback loops to enhance accuracy and domain relevance. Their research highlighted how custom GPTs could streamline the manuscript preparation process by efficiently synthesizing and analyzing scientific literature. However, the study also pointed out persistent inaccuracies, emphasizing the need for continuous model refinement and human oversight to ensure credibility (Kabir et al., 2024). In a similar vein, Gorelik et al. (2024) developed a custom GPT for medical consultations, focusing on the management of pancreatic cysts. By integrating multiple clinical guidelines, their model provided recommendation-based responses that aligned with expert opinions in 87% of cases. The study demonstrated that custom GPTs can achieve expert-level consistency in specific domains, making them valuable tools in decision-making and professional education (Gorelik et al., 2024).

Findings from recent studies suggest that custom GPTs outperform general-purpose AI models in specialized domains. Liu et al. (2024) compared the performance of GPT-3.5, GPT-4, GPT-4o, and custom GPTs on the Emergency Medicine Specialist Examination in Taiwan. Their results showed that custom GPTs achieved higher accuracy than GPT-4 but were still outperformed by GPT-4o. This suggests that while customization improves AI performance, base model capabilities still play a crucial role in determining accuracy and effectiveness (Liu et al., 2024). In the field of education, Kwon (2024) developed a customized GPT chatbot for pre-service teacher training in mathematics. When tested against a general-purpose AI (ChatGPT) and an expert educator, the custom GPT's responses received an average score of 3.73 out of 5, compared to the expert's 4.52 and ChatGPT's 2.86. While the customized chatbot did not fully match expert performance, it demonstrated greater reliability and educational validity than generic AI models, highlighting the potential of context-aware AI for teacher training (Kwon, 2024). However, concerns about privacy, security, and misinformation persist. Antebi et al. (2024) raised concerns about the misuse of custom GPTs, noting that malicious configurations could introduce risks such as data leakage, misinformation propagation, and security vulnerabilities. These concerns underscore the need for robust validation, transparency, and regulatory oversight in the deployment of domain-specific GPTs (Antebi et al., 2024).

**Gaps**

The development and implementation of custom GPTs require continuous refinement, domain-specific training, and responsible oversight to maximize their effectiveness in education and research. Studies emphasize the need for enhanced calibration and accuracy, ensuring AI models undergo feedback-driven improvements and expert validation to generate reliable and contextually accurate responses (Kabir et al., 2024; Gorelik et al., 2024). Additionally, integrating validated, high-quality datasets specific to each field is essential for ensuring that AI-generated content aligns with established knowledge (Liu et al., 2024). Without these refinements, custom GPTs risk producing misleading or overly generalized information, limiting their utility in specialized domains such as teacher education and action research.

Beyond technical improvements, user training, ethical safeguards, and broader accessibility are critical. GeAI-generated content should be accompanied by training programs that help users—particularly PSTs, critically assess AI outputs, verify citations, and apply research findings responsibly (Kwon, 2024). Furthermore, ensuring regulatory oversight and security measures is essential to prevent AI misuse, misinformation, or unauthorized access (Antebi et al., 2024). To make these tools widely accessible, collaborations between AI developers and educational institutions should be expanded, particularly in low-resource regions where AI can bridge learning and research gaps (Gorelik et al., 2024). These recommendations align with efforts to integrate AI into Ghana's teacher education system, reinforcing the importance of localized AI solutions while addressing challenges related to accuracy, security, and accessibility.

## 3. Approach

OpenAI provides functionalities that allow users to customize GPTs to perform specific tasks, making it possible to develop AI models tailored to educational contexts. One of these functionalities includes the Retrieval-Augmented Generation (RAG) approach, which enables the integration of curated datasets, ensuring that the GPT prioritizes reliable, structured, and domain-specific knowledge before exploring external web sources. This approach enhances contextual relevance and accuracy, making AI responses more aligned with educational needs. For this study, we developed customized GPT using these functionalities while implementing an ISA. The ISA ensures that users can interact with the GenAI in a guided manner, reducing the need for advanced prompt engineering skills. The development process involved multiple iterations and extensive curation of educational data to fine-tune the GenAI's responses. The *21st Century Teacher Educator_GH* prioritizes our instructions and preloaded curriculum-based materials and frameworks to provide a contextualized GenAI response to support PSTs in Ghana.

### 3. 1 1st Century Teacher Educator

The *21st Century Teacher Educator_GH* is a context-aware AI-powered tool designed to support student-teachers in Ghana's Initial Teacher Education (ITE) system. By integrating course syllabi, lecture notes, past questions, and curriculum materials, the tool enhances personalized learning and academic support. This tool dynamically interacts with student-teachers. Unlike the base GPT models, the *21st Century Teacher Educator_GH* employs a context-aware approach, allowing it to tailor learning experiences based on students' academic levels, programs, and specific course needs. This section provides a detailed account of the configuration process, explaining how the system was developed to interact dynamically with users, retrieve relevant course materials, and adapt to the specific needs of PSTs.

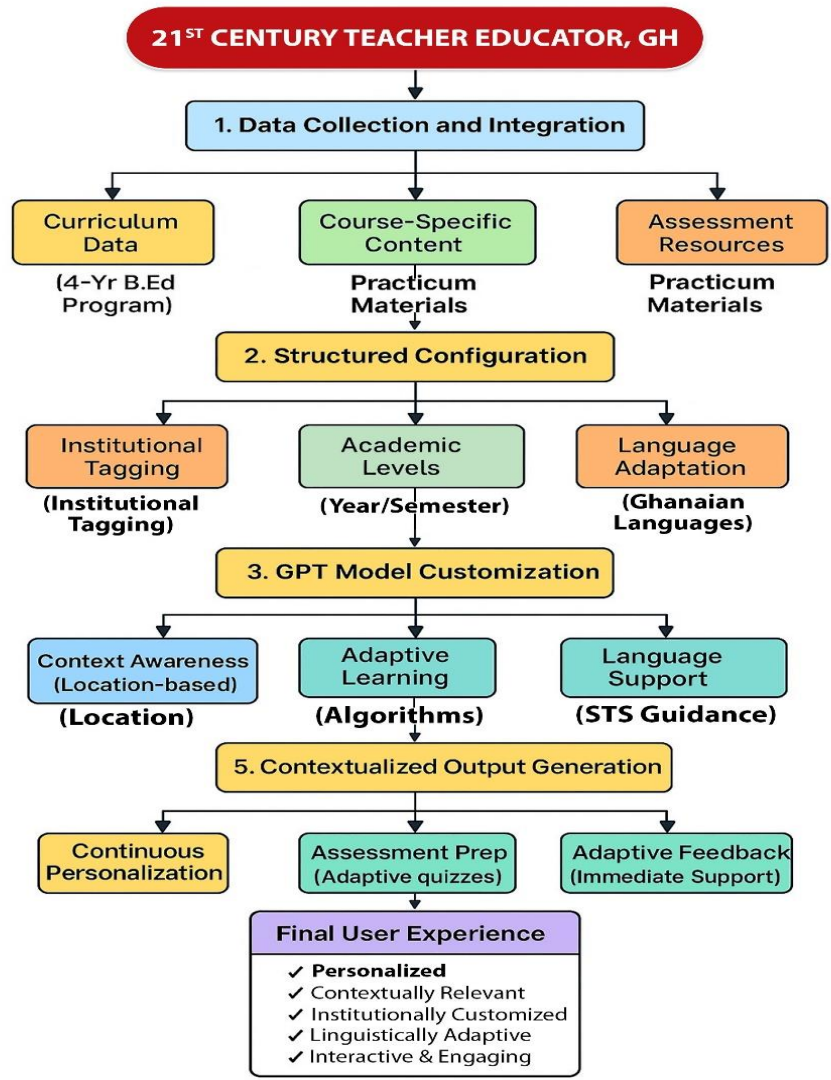

*Figure 2: Architecture of Ghana's 21st Century Teacher Education Customized GPT*

### 3.1.1 Data Integration and Configuration

The development of the *21st Century Teacher Educator_GH* required a robust data integration process to ensure that it could interact meaningfully with student-teachers and provide accurate, curriculum-aligned support. To achieve this, relevant educational resources were carefully collected, categorized, and structured based on Ghana's Initial Teacher Education (ITE) curriculum. This structured approach ensures that when a student-teacher engages with the system, they receive contextually relevant materials tailored to their academic level, program specialization, and current coursework.

At the core of this knowledge system is the Four-Year Bachelor of Education (B.Ed) Curriculum, which serves as the foundational framework for teacher education in Ghana. This document outlines the structured coursework, detailing the various subjects taught over eight semesters, as well as the assessment formats and practicum requirements for student-teachers. By integrating this curriculum into the GPT's knowledge base, the system ensures that students receive information that aligns with national academic standards rather than generic teaching resources.

Beyond the curriculum structure, lecture notes and course outlines were incorporated to provide detailed subject-specific content. These materials are essential in helping student-teachers navigate their enrolled courses, allowing them to access comprehensive explanations of key concepts, theories, and instructional methodologies. Whether a student is studying Inclusive School-Based Enquiry, Early Grade Science, or Differentiated Assessment for Upper Grade, the system retrieves the appropriate notes and presents them in a structured, digestible manner.

To further enhance learning, past questions and assessments were added to the system. Recognizing the importance of practice and self-assessment, exam and quiz questions from previous academic years were integrated, enabling students to test their understanding and prepare effectively for upcoming assessments. The system provides multiple-choice questions, fill-in-the-blank exercises, and essay-style questions, allowing students to engage in different forms of evaluation. Moreover, after answering a set of questions, students have the option to receive immediate feedback and explanations, reinforcing their understanding and identifying areas that need improvement.

An essential component of teacher education in Ghana is Supported Teaching in Schools (STS), which focuses on practical teaching experience. To support student-teachers during their practicums, STS handbooks were incorporated into the system. These documents serve as guides for reflective writing, lesson planning, and action research, helping students effectively document their experiences in real classroom settings. By making these handbooks readily available through the GPT, student-teachers can access step-by-step guidance on teaching methodologies, classroom management strategies, and assessment techniques, all of which are crucial for their professional development.

Through this carefully structured data integration process, the *21st Century Teacher Educator_GH* becomes more than just an AI tool, it transforms into a comprehensive digital companion that supports student-teachers throughout their educational journey. Through the knowledge base with Ghana's official curriculum and academic resources, the system ensures that every interaction is relevant, structured, and responsive to the needs of future educators.

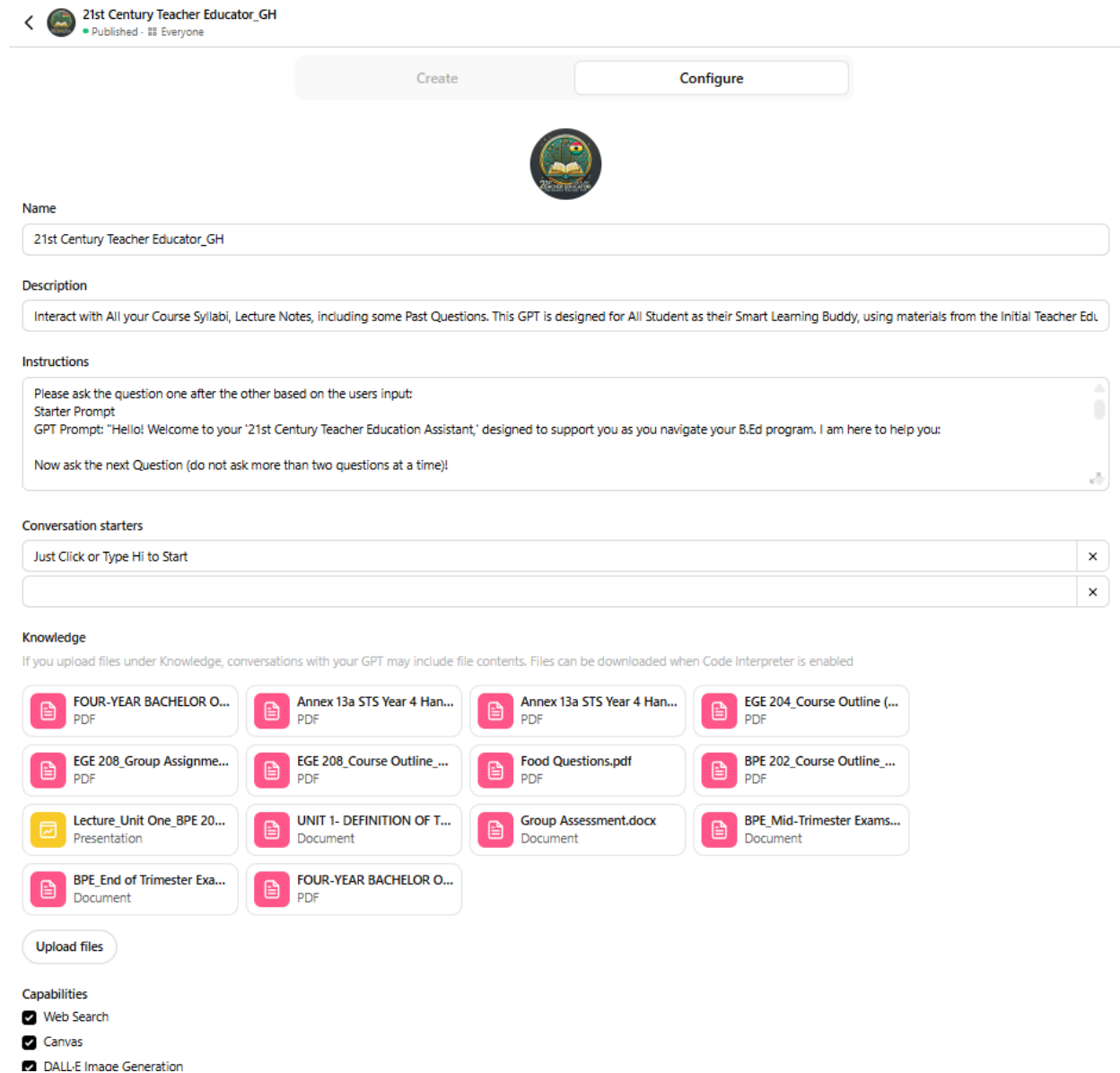

*Figure 3. Data Integration and Configurations*

### 3.1.2 User Interaction and Semi-Auto Prompting

The interaction process within the *21st Century Teacher Educator_GH* is designed to be intuitive and responsive. Upon initiating a session, the system first seeks to understand the student's institutional background. A semi-autoprompt is a structured AI interaction method that provides users with guiding questions, predefined options, or partially completed inputs to streamline their responses while maintaining flexibility. Unlike fully automated prompts that generate responses without user input, or manual prompts where users must type everything from scratch, semi-autoprompts offer a balanced approach following (Nyaaba and Zhai, 2025). This method ensures that interactions remain structured and efficient, especially in educational tools like the *21st Century Teacher Educator_GH*, where students may need assistance navigating large volumes of academic content. By offering dropdown selections, numbered choices, or fill-in-the-blank prompts, the AI makes it easier for users to engage without feeling overwhelmed. For instance, instead of requiring students to manually enter their entire course details, the system might ask, *"Which year are you in? (Type 1, 2, 3, or 4)"* followed by *"Which program? (1. Early Grade, 2. Upper Grade, 3. JHS Specialism)"*, allowing for a more structured and intuitive user experience. reducing the effort required to provide inputs, semi-autoprompts enhance engagement, minimize cognitive load, and improve the overall effectiveness of AI-driven educational support systems.

Given that Ghana has 47 Colleges of Education and multiple universities offering teacher education, this step ensures that the tool can provide institution-specific guidance where necessary. Once the institution is identified, the student is prompted to indicate their academic year and semester, allowing the system to narrow down the list of courses they are currently enrolled in.

With the semester and courses established, the student is then asked to select a specific subject they wish to explore. The GPT responds by presenting a structured overview of the course, including key topics and concepts. If the student requires lecture notes, the system provides summarized explanations alongside real-life examples relevant to the Ghanaian context. In cases where students seek assessment preparation, the system offers multiple-choice questions, fill-in-the-blank exercises, and essay-type questions based on past exams. This feature allows students to test their understanding and even receive feedback on their answers.

Beyond theoretical knowledge, the system plays a crucial role in **teaching practice support**. Students participating in STS can indicate their phase of practicum—whether Beginning Teaching or Embedded Teaching—and receive tailored resources for their reflective practice and lesson documentation. Additionally, if a student requires lesson planning assistance, they are redirected to a specialized culturally responsive lesson planner, ensuring that their instructional designs align with the local educational context.

**4. Use Case: Culture and Context-Awareness**

A defining feature of the *21st Century Teacher Educator_GH* is its ability to adapt dynamically to the diverse culture and context of student-teachers across Ghana (See Figure 1). Recognizing the varied institutional structures, linguistic preferences, and individual learning paths of users, the system goes beyond simply providing static educational content. Instead, it employs a context-aware approach to tailor its responses based on a student's location, institution, language preference, and course requirements. This ensures that the learning experience is not only accurate and relevant but also culturally and linguistically appropriate for the user.

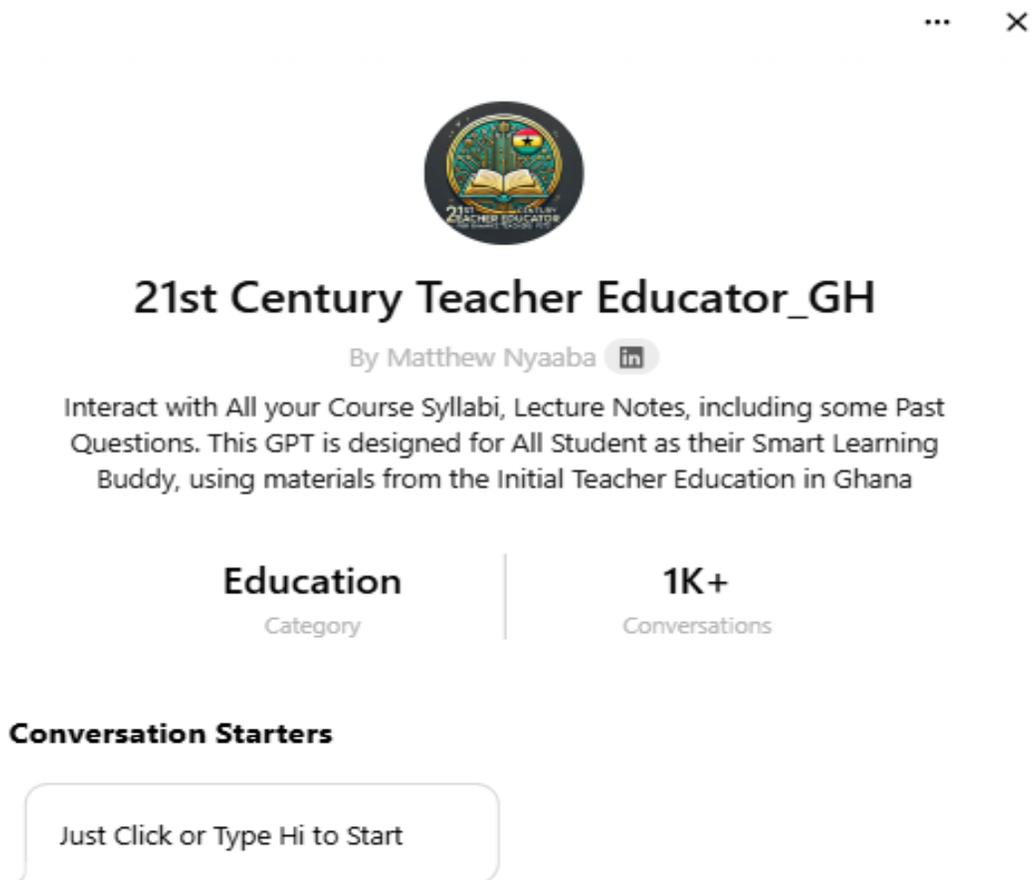

*Figure 4: The 21<sup>st</sup> Century Teacher Educator for Ghana in OpenAI Website ([Link](Link))*

*4.4.1 Institutional Differences: Location-Based Customization*

When a student interacts with the system, one of the first steps is to determine their institution. Ghana has 47 Colleges of Education alongside several universities that offer teacher education programs, and while they all follow the national curriculum, there are subtle differences in the way courses are taught, assessed, and structured across institutions. A student from the University of Cape Coast (UCC), for instance, may have a different assessment format compared to a student from Accra College of Education. Once a student's institution is identified early in the interaction, the tool can situate their learning experience within the context of their specific academic environment.

This differentiation is particularly important for assessment preparation. Suppose a student from a College of Education is preparing for their mid-trimester exams, while a student from a university is working on an end-of-semester assessment. Although both students may be studying the same course, the structure, question format, and evaluation criteria may differ. The system intelligently adapts to these variations, ensuring that each student receives study materials and practice questions that align precisely with their institution's expectations. Beyond assessments, the system's location-based awareness also influences the examples and teaching materials it provides. If a student is studying science education in a rural teacher training college, the tool might suggest examples that incorporate local environmental observations and community-based learning approaches, whereas a student in an urban university setting might receive examples that focus on technology integration in classrooms. This level of contextualization makes learning more meaningful and applicable to the student's future teaching environment.

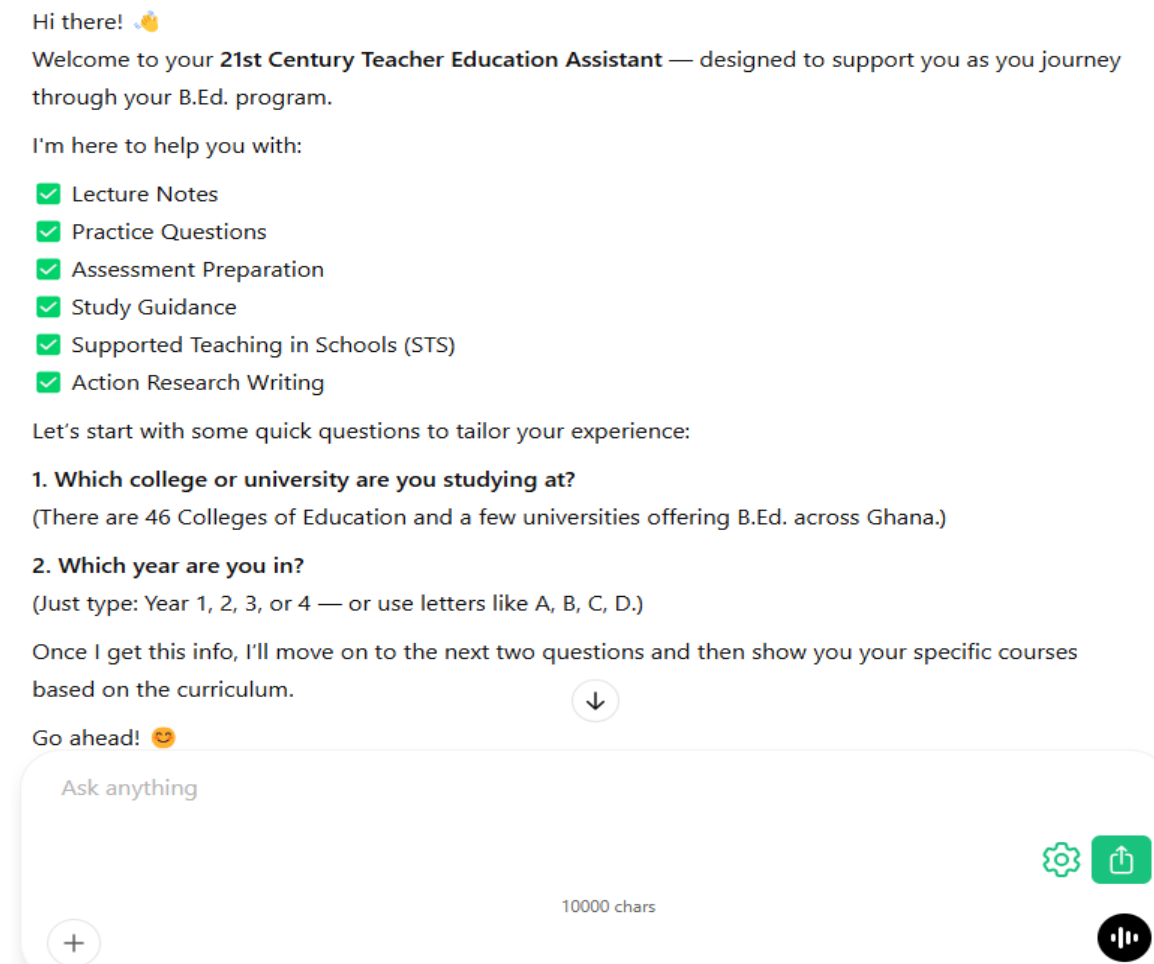

*Figure 5: Context-Based Customization of the 21st Century Teacher Educator for Ghana*

*4.4.2 Language Adaptation and Localization*

In a country as linguistically diverse as Ghana, where English serves as the official language of instruction but numerous indigenous languages play a crucial role in everyday communication, language adaptation is a key component of effective learning. The *21st Century Teacher Educator_GH* recognizes that while students may be proficient in English, some concepts may be better understood when explained in their native language.

To address this, the system is designed to translate and explain educational content in Ghanaian languages such as Gurne, Dagaare, Twi, Ewe, Dagbani basd on th locality chosen or by request by educators. If a student struggles to grasp the concept of inquiry-based learning, for instance, the tool can break it down in Twi as follows:

This feature ensures that even complex pedagogical theories can be made accessible to students who may think more fluently in their first language. Beyond translation, the tool also incorporates culturally relevant examples that reflect the lived experiences of Ghanaian student-teachers. If a student is learning about differentiated assessment strategies, the system might provide an example of how a primary school teacher in Tamale uses oral assessments in Dagbani to support students with limited English proficiency. By embedding local languages and culturally familiar teaching scenarios, the GPT creates an inclusive and relatable learning environment.

## Phonemic Awareness Examples for Students in Gambaga, Ghana

Since Gambaga is in the **North East Region of Ghana**, many children speak **Mampruli** as their first language, with some exposure to **Dagbani, Hausa, and English**. To make phonemic awareness activities relevant and engaging, it's best to use **local words, objects, and cultural experiences** that students can relate to.

### 1 Blending Sounds (Phoneme Blending) – "Say It Fast" Game

**Objective:** Help students blend individual phonemes into complete words.
**Activity:**

- Teacher says the sounds slowly, and students guess the word.
- Use familiar words from Mampruli and English.

**Examples:**

- /k/ /o/ /m/ /b/ /a/ → *komba* (meaning *child* in Mampruli)
- /b/ /u/ /a/ → *bua* (meaning *goat*)
- /m/ /a/ /r/ /k/ /e/ /t/ → *market*
- /t/ /a/ /m/ /a/ → *Tamale* (a city name)

### 2 Segmenting Words – "Chop the Word" Activity

**Objective:** Help students break words into their individual sounds.
**Activity:**

- Teacher says a word, and students break it into phonemes.
- Use local objects they see daily.

*Figure 6: Language Adaption of the 21st Century Teacher Educator for Ghana*

4.4.3 Personalized Course Selection and Learning Pathways

Not all students require the same level of detail in their study materials, and the *21st Century Teacher Educator_GH* is designed to allow users to customize their learning journey. Instead of providing generic lecture notes for an entire course, the system prompts students to select specific topics or subtopics within a course, ensuring that they focus only on what is most relevant to them at any given time.

For example, a student enrolled in EGE 208 (Early Grade Science II) may only need guidance on "Teaching Methods in Science" rather than the entire course content. Instead of overwhelming them with an entire semester's worth of notes, the system retrieves only the relevant section that covers different instructional strategies, such as inquiry-based learning, hands-on experiments, and storytelling in science education. This feature is particularly useful for students who are preparing for exams, writing research papers, or working on specific assignments. By giving students control over their learning pathways, the system fosters self-directed learning and allows students to pace themselves according to their individual needs. Whether they are revisiting foundational concepts or diving into more advanced topics, they can navigate their coursework with precision rather than being burdened with excessive, non-essential information.

🔼 Which semester are you in?
- a) Semester 1
- b) Semester 2

Just type the letter for each response! 😊

> 3. a 4. a

Awesome! You're in **Early Grade (KG – P3), Year 1, Semester 1** at Gambaga College of Education.

Here are the **courses you are studying this semester** based on the curriculum:

1️⃣ Pedagogic Knowledge with ICT & Inclusion: SEN/Gender
2️⃣ Foundations of Education in Ghana
3️⃣ Inclusive School-Based Inquiry
4️⃣ Language and Literacy
5️⃣ Numeracy (Learning, Teaching, and Applying Number and Algebra)
6️⃣ Science (Introduction to Integrated Science I)
7️⃣ Social Studies and TVET (Foundations of Social Studies and TVET)
8️⃣ Supported Teaching in School (STS: Beginning Teaching 1)

- Can you confirm if these are your courses, or do you have any additional ones?

Once confirmed, let me know what you need help with:
📚 Lecture Notes
❓ Practice Questions (Quiz, Midterm, or Exams)
📝 Study Guidance
🎓 Teaching Practice (STS support)

Just type the number or request! 😊

*Figure 7: Personalized Course Selection and Learning Pathways of the 21st Century Teacher Educator for Ghana*

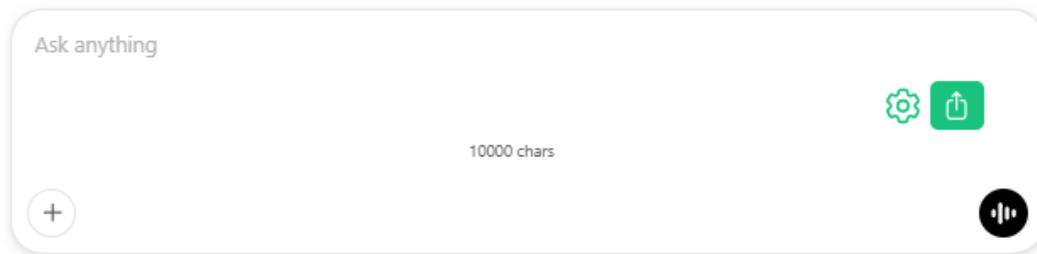

*Figure 6: Course-specific Lecture Note of the 21$^{st}$ Century Teacher Educator for Ghana*

1.4.4 Practice Questions for Course Mastery

In addition to offering differentiated instruction and localized content delivery, the 21st Century Teacher Educator_GH system also leverages GPT capabilities to generate practice questions and past exam-style items to reinforce students' understanding of course content. This function allows pre-service teachers to self-assess their knowledge, identify learning gaps, and build confidence ahead of formal evaluations. For example, in the *Inclusive School-Based Inquiry* course, the system automatically generated multiple-choice questions aligned with specific course objectives, such as identifying strategies for inclusive classroom practices or recognizing barriers to participation among learners with disabilities. These questions were generated based on curriculum-aligned tags and student-selected subtopics, offering contextually relevant and accurate assessment support. By integrating such practice opportunities, the tool not only promotes active learning and retrieval practice but also ensures that learners with different cognitive abilities, literacy levels, and learning styles, as outlined in the key learner characteristics, can engage with the content in ways that suit their individual needs. Ultimately, this function supports the development of reflective, responsive, and well-prepared future educators.

## Practice Questions on Learner Characteristics

(For Inclusive School-Based Enquiry)

**Section A: Multiple Choice Questions (MCQs) – 10 Marks**

1. Which of the following is NOT a type of learning style?
a) Visual
b) Auditory
c) Reflective
d) Kinesthetic

2. A child who learns best through hands-on activities and movement is a:
a) Visual learner
b) Auditory learner
c) Kinesthetic learner
d) Reading/Writing learner

3. What is the best way to support a student who has difficulty hearing?
a) Speak louder in class
b) Use sign language or visual aids
c) Ask them to sit at the back of the class
d) Ignore them and continue the lesson

4. A child from a farming background may struggle with reading but excel in:
a) Oral storytelling
b) Writing essays
c) Hands-on activities
d) Silent reading

5. Which factor affects learners' ability to communicate in class?

*Figure 7: Practice Questions for the 21$^{st}$ Century Teacher Education for Ghana*

**5 Launch and Stakeholders' Feedback**

On Saturday, March 8, 2025, Bagabaga College of Education (BACE), in collaboration with the National Union of Ghana Students (NUGS) and the Generative Artificial Intelligence for Education and Research in Africa (GenAI-ERA), hosted a landmark training event themed "The Effective Use of Artificial Intelligence in Our Educational Settings." This event marked the official launch of a broader initiative aimed at transforming teacher education through the ethical and responsible integration of Generative AI (GenAI) tools. The training, which drew approximately 250 students and several faculty members, was chaired by the Vice Principal of BACE. Among the key innovations unveiled was the 21st Century Smart Learning Buddy. This tool was introduced as a strategic response to curriculum delivery challenges, particularly supporting pre-service teachers and lecturers in automating content navigation, instructional planning, and pedagogical coherence within the curriculum framework.

Both teacher educators and PSTs highlighted the 21st Century Teacher Educator_Ghana's relevance in addressing gaps in instructional support and praised its contextual design tailored to local educational realities. Teacher educators noted the tool's potential in enhancing the efficiency and quality of teacher preparation, particularly in ensuring consistent alignment with national curriculum goals. The interactive demonstration led by GenAI-ERA co-founder showed how the tool could be practically integrated into pre-service teachers' studies and teaching methodologies. Further discussion from the PSTs also revealed a strong interest in scaling up the 21st Century Teacher Educator to other teacher education institutions across the country. Feedback consistently pointed to the Smart Learning Buddy as more than a technological aid; it was perceived as a transformative catalyst for 21st-century teacher education, bridging innovation, equity, and localized pedagogical practice.

**6. Discussion**

The design of the 21st Century Teacher Educator_GH GPT is shaped by a constellation of interrelated frameworks that prioritize ethical GenAI integration, cultural responsiveness, decolonial design principles, and fidelity to national curriculum standards. These frameworks, drawn from both global and local sources, work together to ensure the GPT is a tool of convenience, and a transformative learning companion for PSTs in Ghana.

At the global level, this project draws upon the UNESCO (2023) framework on co-designing GenAI to support teachers and teaching, which emphasizes that AI must serve as a supportive agent, not a substitute, for teacher agency and decision-making. The GPT was intentionally built to preserve this agency, which ensures that PSTs remain in control of their learning pathways. It invites them to choose courses, define learning goals, and explore content relevant to their immediate environments. In doing so, the GPT honors UNESCO's call to place educators at the center of GenAI design and promotes pedagogical interactions that are both responsible and human-led.

The GPT that adapts to the user's geographic and cultural background. For example, a PST in the Northern Region receives science content embedded with *Mampruli t*erms and community-based analogies, while a student in the Ashanti Region is supported with *Twi* translations and localized teaching examples. These interactions are not just personalized but they are culturally grounded. This aligns the culture and context-aware framework (Nyaaba and Zhai, 2025) as well as study by Kabir et al. (2024) and Gorelik et al. (2024) which highlights how customized GPTs enhance accuracy and relevance in specialized fields, a claim reinforced by the 21st Century Teacher Educator, which provide structured, to pre-service teachers (PSTs). This framework was specifically developed to guide GPT customization for culturally diverse and data-marginalized contexts. In this framework is the Gay's (2000) work on cultural competence and Ladson-Billings' (1994) foundational principles of culturally relevant pedagogy, the framework emphasizes contextual accuracy, linguistic inclusivity, and cultural affirmations (Nyaaba and Zhai, 2025). Moreover, the GPT adapted into the generates culturally sensitive self-assessments, provide place-based instructional support, and scaffold action research in a way that reflects local ways of knowing and learning. This aligns with GenAI-CRSciA framework that introduces tenets such as multilingual inclusion, indigenous knowledge recognition, and community and family engagement (Nyaaba et al., 2024).

Adding a critical and liberatory lens, the design also draws heavily on the framework for mitigating GenAI neocolonialism (Nyaaba et al., 2024). This framework calls for an intentional shift away from dominant Western-centric AI infrastructures and proposes strategies to ensure more equitable GenAI development for the Global South. Key to this approach is the application of LDM and FBD, approaches that empower designers to engage with marginalized identities, anticipate harm, and center local narratives as sources of innovation. In the GPT, these ideas take form through prompt contextualization, where PSTs are encouraged to craft culturally grounded inquiries that resonate with the PSTs local knowledge and context dismantle the one-size-fits-all approach that has historically marginalized African educational contexts in global design. In line with the Ghana's NTECF, upon logging in, PSTs are prompted to enter their year, semester, and institution, allowing the GPT to identify their current coursework (MoE, 2018). It then generates content aligned with the national curriculum, ranging from lecture notes and examples to self-assessment questions and practicum support, all grounded in Ghanaian contexts and values.

One of the key challenges observed during implementation was the presence of hallucinations, particularly in instances where certain Ghanaian languages were either underrepresented or not adequately captured by mainstream language models. This limitation is likely since some local languages in Ghana have not been fully transcribed into written form or lack sufficient digital corpora. This supports the concerns raised by Antebi et al. (2024) regarding misinformation, security risks, and ethical vulnerabilities in customized GPTs remain relevant. Future research should explore strategies for sustained integration, institutional collaboration, and regulatory oversight, ensuring that GenAI remains a complementary, rather than a replacement, for human-driven teacher education.

**Conclusion and Future Research**

The development of the 21st Century Teacher Educator_GH system was guided by the Glocalized GenAI in Education Framework, which integrates six critical strands to ensure contextually relevant, ethical, and inclusive AI use in teacher education. Drawing from the UNESCO Framework, the system prioritizes educator autonomy and human agency, empowering pre-service teachers to take ownership of their learning. The emphasis on Mitigating Neocolonialism informed the use of liberatory design and decentralized AI development to center African educational needs and realities. Cultural relevance was further enhanced through the Culture and Context-Aware Framework, which ensured the integration of local examples and culturally responsive pedagogies. The GenAI-CRSciA strand supported multilingual strategies and the inclusion of Indigenous knowledge, while the National and Disciplinary Curriculum Frameworks guided the alignment of content with Ghana's teacher education structure.

These frameworks collectively informed the technical and pedagogical architecture of the system, making it both globally informed and locally grounded.

Pilot testing of the platform highlighted its effectiveness in promoting language adaptation, localization, and personalized learning pathways. By enabling the translation of key educational content into Ghanaian languages such as Twi, Ewe, Dagbani, Gurne, and Dagaare, the system helped break linguistic barriers and ensured deeper conceptual understanding for students. Culturally familiar examples made complex pedagogical ideas, like differentiated instruction or inquiry-based learning, more accessible and meaningful. Additionally, the tool allowed students to customize their learning journeys by selecting specific topics or subtopics rather than navigating entire course content, fostering self-directed learning. These features, grounded in the six foundational frameworks, position the 21st Century Teacher Educator_GH as a transformative innovation with potential to be scaled across other linguistically and culturally diverse teacher education contexts in Africa. However, subscription-based limitations restrict full utilization by PSTs and teacher educators. To address this, collaboration with OpenAI's NextGen initiative could expand EduGPT access across Ghana's teacher training colleges. Future research should assess long-term impacts on academic performance and teaching effectiveness to determine the full potentials in shaping GenAI-enhanced teacher education.